\documentclass[twocolumn,aps,nofootinbib]{revtex4-2}
\usepackage{graphicx}
\usepackage{comment}
\usepackage{enumitem}
\usepackage{xcolor}
\usepackage{soul}
\usepackage[breaklinks]{hyperref}
\usepackage{url}

\usepackage{breakurl}

\usepackage{multirow}
\usepackage{setspace}
\usepackage{amsmath, amssymb, amsthm}
\usepackage{cleveref}
\usepackage{physics}

\usepackage{crimson}

\color[rgb]{0,0,0} %black

\begin{document}

\title{Modeling complex systems: A case study of compartmental models in epidemiology}

\author{Alexander F. Siegenfeld$^{1,2}$}\thanks{These two authors contributed equally.}

\author{Pratyush K. Kollepara$^{1,3}$}\thanks{These two authors contributed equally.}

\author{Yaneer Bar-Yam$^{1}$}

\affiliation{$^1$New England Complex Systems Institute, Cambridge, MA, USA}
\affiliation{$^2$Department of Physics, Massachusetts Institute of Technology, Cambridge, MA, USA}
\affiliation{$^3$Department of Mathematical and Physical Sciences, La Trobe University, Melbourne, Australia}

%\date{\small \today}

\begin{abstract}    
Compartmental epidemic models have been widely used for predicting the course of epidemics, from estimating the basic reproduction number to guiding intervention policies.  Studies commonly acknowledge these models' assumptions but less often justify their validity in the specific context in which they are being used.  Our purpose is not to argue for specific alternatives or modifications to compartmental models, but rather to show how assumptions can constrain model outcomes to a narrow portion of the wide landscape of potential epidemic behaviors. This concrete examination of well-known models also serves to illustrate general principles of modeling that can be applied in other contexts.
\end{abstract}
\maketitle

\section{Introduction}
Compartmental models such as the SIR model have been widely used to study infectious disease outbreaks~\cite{Hethcote2000, Rock2014, Barrat2008, PastorSatorras2015, Billah_2020}. These models have informed policy makers of the risks of inaction and have been used to analyze various policy responses.  The limitations of the assumptions of compartmental models are well-known \cite{Tolles2020, Wolfram_SIR, Roberts2015, Givan2011, DHAR2020}; we intend to explore which assumptions are appropriate in which contexts and when and why the models do or do not succeed.
%however, such models are often applied without a careful analysis of which assumptions are appropriate in which contexts. 

No model accurately captures all the details of the system that it represents, but some models are nonetheless accurate because certain large-scale behaviors of systems do not depend on all these details~\cite{Siegenfeld2020}.  (For example, modeling material phase transitions generally does not require including the quantum mechanical details of individual atoms.)  The key to good modeling is understanding which details matter and which do not.
Paradoxically, failing to recognize that a model can be accurate in spite of certain unrealistic assumptions can lead to models in which all assumptions are excused: the impossibility of getting all the details right may discourage a careful analysis of which assumptions are appropriate in which contexts.

\begin{figure*}%[t]
    \centering
    \includegraphics[scale=0.4, trim={0 0 0 0}, clip]{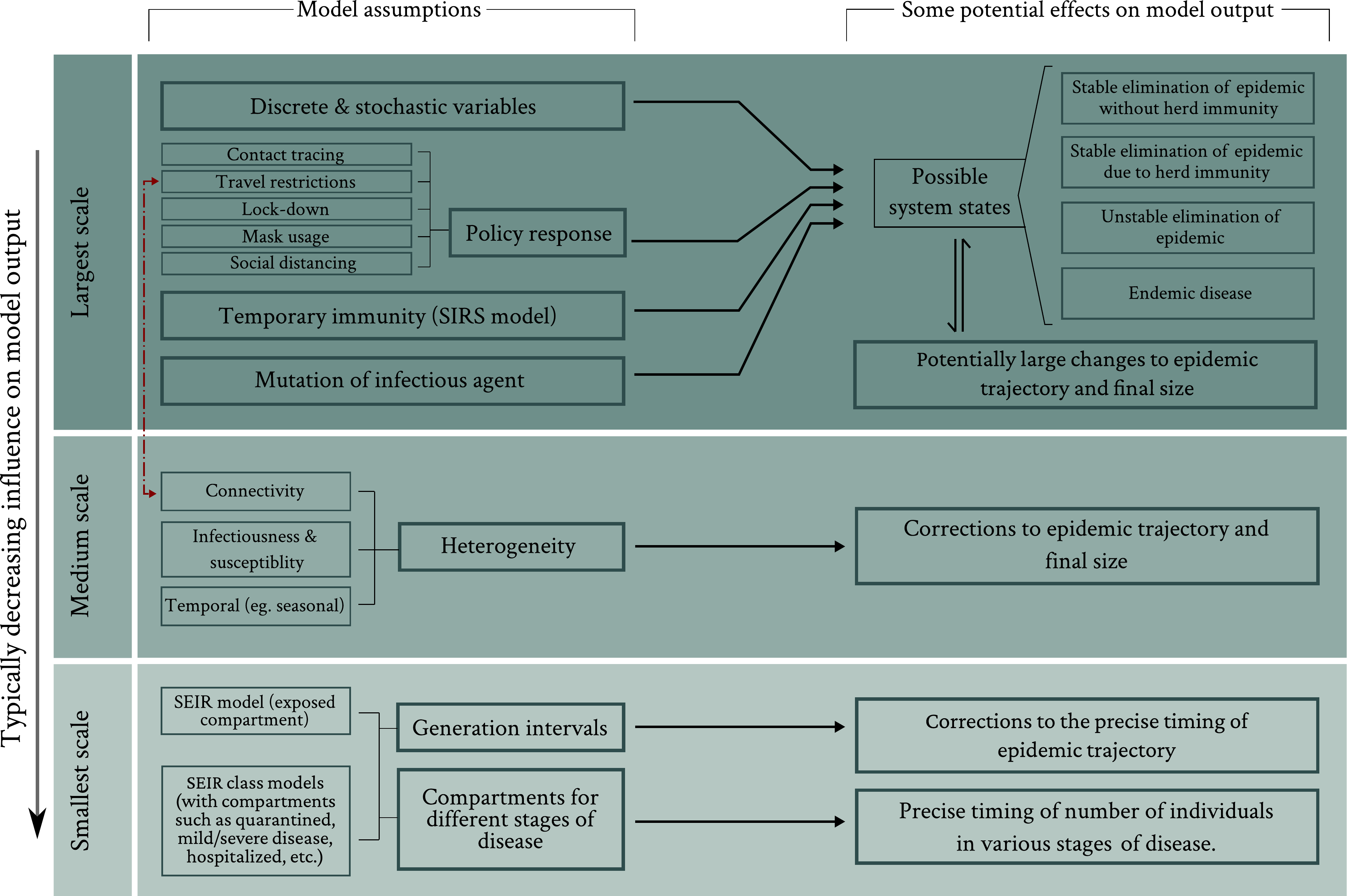}
    \caption{Schematic representation of the impact of various modeling choices/assumptions. The left column lists various details that can be incorporated into a compartmental model, and the right column lists typical potential impacts on the model output.    The three panels classify the details by `scale', with the largest scale details typically having the most impact on model output, and the smallest scale details typically having the least impact, although the impact of any given assumption ultimately depends on precisely for what purpose the model is being used.  %The details listed in the left column can compound and cause non-linear changes in the model output. %For instance, discrete and stochastic variables by themselves are not sufficient to maintain stable elimination, but appropriate policy response like contact tracing are required to ensure that cases imported from elsewhere do not lead to a new outbreak or the mutation of infectious agents can be influenced by selection pressure from policy interventions.
    %Within each colored panel, it is difficult to determine which details have the most impact.
    %The model objectives should be considered for assessing the importance of the details. 
Furthermore, various assumptions can compound non-linearly to affect the model output. For instance, policy interventions such as travel restrictions, which both rely on and affect heterogeneity in geographic connectivity, can play a decisive role in determining whether or not a stable elimination is achieved~\cite{Siegenfeld2020_Commphys}. %Further, if geographical connectivity is not considered in the model, then the model will not be able to recognize elimination in some of the local regimes and may mis-characterize the system state (See Section \ref{stochastic}).
    Of course, the actual effect of any assumption depends on its precise mathematical implementation, as well as the presence or absence of other assumptions within the model, and so this figure should be considered as a rough schematic rather than as a definitive guide.}
    %The details listed in the left column can compound and cause non-linear changes in the model output. %For instance, discrete and stochastic variables by themselves are not sufficient to maintain stable elimination, but appropriate policy response like contact tracing are required to ensure that cases imported from elsewhere do not lead to a new outbreak or the mutation of infectious agents can be influenced by selection pressure from policy interventions.
    %Within each colored panel, it is difficult to determine which details have the most impact.
    %The model objectives should be considered for assessing the importance of the details. 
    
    %Note that a detail of smaller scale does not imply that it is a less useful detail. For example using models with compartments like mild or severe disease may be helpful for designing an appropriate policy response, but including such details while ignoring other large scale details (which have a far larger impact on the output) is incorrect.
    \label{fig:diagram}
\end{figure*}

%During a pandemic, it is crucial that models complement decision-making. 
In an attempt to obtain better predictions, it may be tempting to include more details and fine-tune the model assumptions. %But an epidemic is a complex system, which depends on many details ranging from biological characteristics of the infectious agent to complicated and unpredictable human collective behavior, and
However, arbitrarily focusing on some assumptions and details while losing sight of others is counterproductive~\cite{Siegenfeld2020_PNAS}. Which details are relevant depends on the question at hand; the inclusion or exclusion of details in a model must be justified depending on the modeling objectives. Compartmental models tend to include some details (e.g. disease stages) while not including others (e.g. stochasticity and heterogeneity) that, in many cases, have a far larger effect on forecasting the epidemic trajectory, estimating the final epidemic size, and analyzing the impact of  interventions (see Figure \ref{fig:diagram}).

%Compartmental models tend to include some details (e.g. disease stages) while not including others (e.g. heterogeneity) that can have a far larger effect on the overall output of the model. 

Sensitivity analysis is becoming a common method for assessing how uncertainties in model parameters can lead to uncertainty in the results.  However, sensitivity analyses can only capture sources of uncertainty that arise from the parameters or assumptions that are explicitly varied, which will often not span entire space of possible epidemic behaviors.  For instance, accounting for uncertainty in the transmission rate of a model that assumes a well-mixed population will not ameliorate errors arising from this latter assumption. 

In this work, we use both theoretical analyses and numerical simulations to examine some common assumptions of compartmental models---such as the distribution of generation intervals, homogeneity in population characteristics and connectivity, and the use of continuous variables---in order to determine their relevance for various model outcomes. Our purpose is not to argue for specific alternatives to compartmental models or for specific modifications but rather to illustrate how the assumptions of these models affect their results. 
\vspace{-3em}
%Indeed, even the heterogeneous compartmental models we explore, still make many assumptions that may lead to an inaccurate characterization of large-scale behavior.  
%In many cases, there is just too much uncertainty to accurately predict pandemic outcomes and models should instead be used to guide action under uncertainty \cite{Siegenfeld2020_PNAS,Diekmann1990, ball1997, Siegenfeld2020_Commphys, Li2020, Maier2020, Althouse2020,  Mancastroppa2021}.

\section{The SIR model}
We first review the basic SIR model. The model divides the population into three compartments---the fractions of individuals who are susceptible ($s$), infectious ($i$), and recovered ($r$). A set of three differential equations governs the dynamics:
\begin{align} \label{eqn:SIR}
    \dv{s}{t} &= - \beta s i \\
    \dv{i}{t} &= \beta s i - \gamma i \\
    \dv{r}{t} &= \gamma i 
\end{align}
The parameter $\beta > 0$ represents the rate at which infectious individuals transmit the disease. Infectious individuals become no longer infectious (recovered or removed) at a rate $\gamma>0$. 
Assumptions of the SIR model include homogeneity in the infectiousness, susceptibility, and connectivity of the population, exponentially distributed recovery times and generation intervals, that discrete and stochastic dynamics can be approximated with continuous and deterministic variables, and that there are no changes over time in the behaviors of either the population or the infectious agent. 
%Constant infectiousness, homogeneity and exponentially distributed recovery times also imply that generation intervals have the same distribution as the recovery times. 
%This model and its variations (SEIR class models) are commonly used to forecast the trajectory of the epidemic and determine the basic reproduction number for emerging outbreaks.
%Given all of these assumptions, it may be surprising that the SIR model works at all.  But it's important to understand that not all of the assumptions matter, i.e. not all of them affect the large-scale behavior of the model.  

By recasting the equations of the model in terms of the basic reproduction number $R_0=\beta/\gamma$,
\begin{align}\label{eqn:SIR_alt}
    \dv{s}{(\gamma t)} &= - R_0 s i \\
    \dv{i}{(\gamma t)} &= (R_0 s - 1) i \\
    \dv{r}{(\gamma t)} &=  i 
\end{align}
it can be seen that the evolution of the system state (i.e. the fraction of people in each of the three compartments) depends only on $R_0$ and that $\gamma$ sets the time scale for this evolution (i.e. a change in $\gamma$ would correspond simply to a stretching or compression of the time axis). Indeed, it can be proven that the final size of an epidemic depends only on the network of probabilities of individuals infecting each other and not at all on how quickly individuals recover or any other time-scales associated with the progression of the disease within individuals \cite{Miller2012, Andreasen2011, Ma2006, ball1997, barbour1990} (see Appendix~section~\ref{app:size} for more details).

This overall time-scale of the epidemic (set by $\gamma$ in the above formulation) is an important parameter; for instance, together with $R_0$, it tells us how quickly case counts will grow.   The SIR model describes this overall time-scale without the need for any additional compartments or parameters.   Additional compartments 
%that have focused on the details of infectious periods, latent periods, and other disease stages 
can provide more information as to the precise timing (as opposed to simply the overall fraction) of the number of individuals in particular disease stages (e.g. exposed, infectious, hospitalized, etc.) and can help us understand, for instance, lags between infections and hospitalizations.  However, such details will often have much smaller effects than regularly used assumptions that impact the overall epidemic trajectory, such as homogeneity, mean-field connectivity, and continuous variables (see section \ref{assumptions} below). 

%Refining the less important assumptions while ignoring the more important ones increases only the appearance of model sophistication. %are added to compartmental models  Adding more details to models, it can lead to a false sense of confidence in the model. %The model is indeed more detailed, but whether the details are important or not, should also be considered. 

\vspace{-.5em}
\section{Analysis of key assumptions} \label{assumptions}
\vspace{-.5em}
We now examine some key assumptions of compartmental models. In section \ref{GI}, we show that assumptions concerning the distribution of generation intervals (i.e. assumptions about diseases stages, recovery rates, etc.) can be captured by an SIR model with appropriately selected parameters. However, in contrast to how the distribution of generation intervals can be described by the effective parameter $\gamma$,  heterogeneous susceptibility and connectivity cannot be captured by an effective spreading rate $\beta$ (section \ref{het}).  In section \ref{stochastic}, we discuss the implications of using continuous and deterministic variables to describe dynamics that are in reality stochastic and discrete. 
\vspace{-.5em}
\subsection{Generation intervals and effective parameters} \label{GI}
\vspace{-.5em}
There are some cases in which simplifying assumptions are not critical.  For instance, by assuming a constant recovery rate $\gamma$, the SIR model makes the assumption that generation intervals follow an exponential distribution.  However, the growth rate and reproduction number can nonetheless be accurately captured despite the actual generation intervals not being exponentially distributed, so long as $\gamma$ is treated as an effective parameter.  

A general result relating the exponent of growth or decline $\lambda$ and the effective reproduction number $R$ is
\begin{equation}
\label{eqn:Wallinga}
    \frac{1}{R}=\int_0^\infty g(t)e^{-\lambda t}dt
\end{equation}
where $g(t)$ is the distribution of generation intervals~\cite{Wallinga2006}. This relationship applies whenever the population size and number of infections is large enough that stochastic effects can be ignored.  (There is also the implicit assumption that $R$ and the generation interval distribution are roughly constant over the time interval during which exponential growth/decline is observed.) 

Within an SIR model, the generation intervals are exponentially distributed with mean $1/\gamma$, so equation (\ref{eqn:Wallinga}) yields 
\begin{align} \label{eqn:Wallinga_SIR}
R_0 = 1 + \lambda_0/\gamma % = 1 + \lambda \mathbb{E}[T] 
\end{align}
where $\lambda_0$ is the initial exponential growth rate. Thus, if an SIR model is to accurately describe $R_0$ and $\lambda_0$ for an observed epidemic, then $\gamma$ must be determined by equation~(\ref{eqn:Wallinga_SIR}). But since actual generation intervals are not exponentially distributed, the inverse recovery rate $1/\gamma$ cannot be estimated as the mean of the observed generation intervals. Instead, $\gamma$ (and $\beta$) serve as effective parameters that coarse-grain the actual generation interval distribution $g(t)$ in such a way that the SIR model yields the correct initial growth rate $\lambda_0$ and basic reproduction number $R_0$:
\begin{align}
    \gamma &= \frac{\lambda_0}{R_0-1}=\frac{\lambda_0}{1/\int_0^\infty g(t)e^{-\lambda_0 t}dt-1} \neq 1/\bar g  \\
    \beta&=\gamma R_0 = \frac{\lambda_0}{1-\int_0^\infty g(t)e^{-\lambda_0 t}dt}
\end{align} 
  where $\bar g$ is the mean of $g(t)$.  Nonetheless, much of the modeling literature (e.g.~\cite{Choi2020, Hyafil2020, Kuniya2020, Read2020, Tang2020, Wu2020, Zhou2020, Kyrychko2020, Walker2020, Davies2020, Chowdhury2020, Radulescu2020, Scala2020, Balabdaoui2020, Balcan2009, Balcan2010, Chinazzi2020, DiDomenico2020, Ferguson2020}) uses $\gamma=1/\bar g$.  This same logic applies to other compartmental models as well: unless there is reason to believe that the time intervals spent between compartments actually follow the distributions implied by the model, the transition rates should be considered as effective parameters.  
\iffalse
Given an observed reproduction number $R_0$ and initial exponential growth rate $\lambda_0$, one may always find a $\beta$ and $\gamma$ such that $R_0=\beta/\gamma$ and $\lambda_0=\beta-\gamma$.  We note in this case that $1/\gamma$ may not be the mean of the distribution of generation intervals $g(t)$.  Instead, the relationship between $\gamma$, $\lambda_0$, and $R_0$ is given by  
\begin{equation}
    \frac{1}{R_0}=\int_0^\infty g(t)e^{-\lambda_0 t}dt
\end{equation}
which implies that the $\gamma$ necessary to match $R_0$ and $\lambda_0$ will be the inverse mean of $g(t)$ if and only if $g(t)=\gamma e^{-\gamma t}$ \cite{Wallinga2006} (see Appendix~\ref{app:GI} for details).
\fi

\begin{figure}%[b]
    \centering
    \includegraphics[scale=1.0, trim={0 0 0 0}, clip]{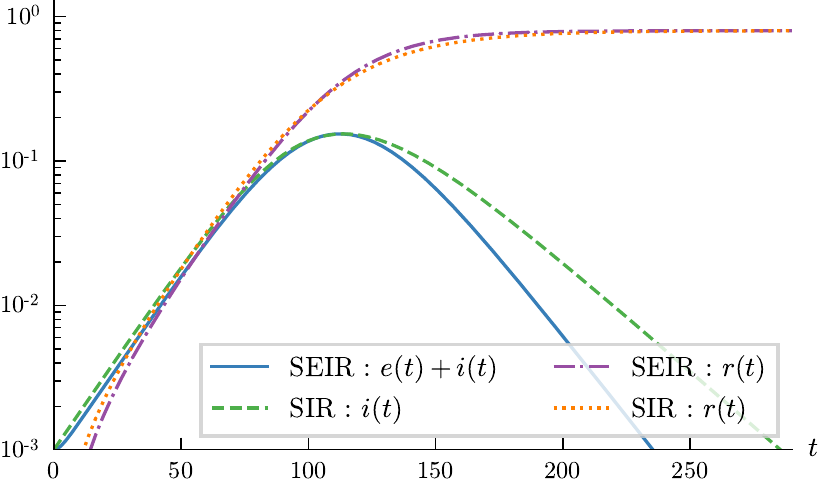}
    \caption{An SEIR model can be replaced with an SIR model with a nearly identical trajectory. Important characteristics such as the growth rate, reproduction number, herd immunity threshold, and epidemic size will be exactly the same in both models.
    The following parameters were used to generate this figure. SEIR: $\beta = 0.25$, $\sigma = 0.167$, $\alpha = 0.125$. SIR: $ \beta = 0.119$, $\gamma = 0.059$ }
    \label{fig:SIR_SEIR}
\end{figure}

Models with additional compartments such as SEIR models are often considered to be more accurate than the SIR model since they include a more realistic effective distribution of generation intervals (see Appendix~\ref{app:SEIR}). However, the precise generation interval distribution does not affect epidemic characteristics such as the final size, the initial exponential growth rate, and $R_0$~\cite{Newman2002}. As described above, these characteristics can be captured by the SIR model by treating the recovery rate $\gamma$ as an effective parameter (see Figure \ref{fig:SIR_SEIR}). More generally, for the purposes of modelling the overall epidemic trajectory, introducing any number of disease stages into the SIR model only amounts to changing the effective distribution of generation intervals, which changes only the timing of the epidemic curve (see Appendix~\ref{app:size}). %The SIR model is elegant in that its parameter set (the dimensionless $R_0$ plus a time scale $\gamma$) is minimal; 
Given the larger sources of uncertainty related to other assumptions, additional parameters in SEIR models are not justified if they serve only to refine the generation interval distribution. (The use of SEIR models over the SIR model may be justified in other circumstances.)

%SIR, SEIR and other compartmental models are frequently used for estimating the basic reproduction number \cite{Billah_2020}, but very often, the transition rates are not treated as effective parameters and are estimated using the inverse of mean generation, serial, latent or infectious intervals \cite{Choi2020, Hyafil2020, Kuniya2020, Read2020, Tang2020, Wu2020, Zhou2020}, which, as described above, is appropriate only for exponentially distributed intervals.  This practice is also prevalent in other epidemic modeling literature \cite{Kyrychko2020, Walker2020, Davies2020, Chowdhury2020, Radulescu2020, Scala2020, Balabdaoui2020, Balcan2009, Balcan2010, Chinazzi2020, DiDomenico2020, Ferguson2020}. 

\vspace{-.5em}
\subsection{Population heterogeneity} \label{het}
\vspace{-.5em}
%Depending on the scale at which an epidemic is being studied, the population can be heterogeneous in one or more ways, such as: individuals in an urban area may have a larger number of contacts than individuals in a rural area, some individuals may be more infectious or susceptible than others (depending on their age, health condition, and mask wearing frequency), individuals may be more likely to travel to nearby regions than far-away regions, or different regions may have different mitigation responses, leading to heterogeneity on the basis of geography. Thus, homogeneity is a key assumption of compartmental models which has been used widely for modeling the coronavirus epidemics \cite{Kyrychko2020, Walker2020, Davies2020, Chowdhury2020, Ferguson2020} and has also been examined in the epidemiology literature by introducing heterogeneity in various ways \cite{Britton2020, Miller2012, Gou2017, Gerasimov2021, Hickson2014, Dolbeault2021}. These analyses of heterogeneity suggest a crucial result that homogeneous models can not effectively coarse grain the heterogeneity in the system.

Human populations are heterogeneous in many ways:  social networks of individuals exhibit community structure ~\cite{Girvan2002, arenas2004, Hedayatifar2019}, infectiousness and susceptibility can vary across the population depending upon age/health/behavior, different regions may have different mitigation responses to an epidemic, etc. The widely used assumption of a homogeneous and well-mixed population has been challenged using various types of heterogeneous models \cite{Britton2020, Miller2012, Gou2017, Gerasimov2021, Hickson2014, Dolbeault2021, Diekmann1990, hebert2020, lloyd2005, woolhouse1997}. %The effects of the homogeneity have been extensively studied using various kinds of heterogeneous models \cite{Britton2020, Miller2012, Gou2017, Gerasimov2021, Hickson2014, Dolbeault2021} and these studies point towards a crucial result: homogeneous models can not effectively coarse-grain the heterogeneity of human populations.
%To demonstrate the impact of heterogeneous infectiousness, susceptibility and connectivity, we use a simple model that presents the previous work in a pedagogical manner. The modeling framework partitions the population into multiple groups
To summarize the impact of heterogeneous infectiousness, susceptibility, and connectivity, we use a simple class of models in which the population is partitioned into multiple groups \cite{Britton2020, Miller2012, Gou2017, Gerasimov2021, Hickson2014, Dolbeault2021}.  The purpose of these modifications is to show that heterogeneity can have a substantial effect on both the epidemic trajectory and its final size and, crucially, can greatly expand the space of possible policy responses.  We do not claim that any particular set of assumptions will accurately predict an epidemic trajectory but rather include these heterogeneous models to show that the space of possible outcomes is far larger than homogeneous models would imply. 

We use a framework in which a population of size $N$ is divided into multiple groups. Each group $k$ has $N_k$ individuals in total, and we define $n_k \equiv N_k/N$ as the fraction of individuals in group $k$, such that $\sum_k n_k=1$. The number of susceptible, infected and recovered individuals in group $k$ is $S_k, I_k, R_k$, respectively, with $s_k \equiv S_k/N, i_k \equiv I_k/N, r_k \equiv R_k/N$, such that $s_k + i_k + r_k = n_k$.  The modified SIR equations can be written as: 
\begin{align}
  \dv{s_k}{t} &= - s_k \sum_{l} B_{kl} i_l \label{eqn:s_eqn}\\ 
    \dv{i_k}{t} &= s_k \sum_{l} B_{kl} i_l - \gamma_k i_k \label{eqn:i_eqn}\\ 
    \dv{r_k}{t} &= \gamma_k i_k \label{eqn:r_eqn}
\end{align}
where $B_{kl}$ represents the rate of transmission from infectious individuals in group $l$ to susceptible individuals in group $k$.  An SIR model is used due to its simplicity, but the key results of this section apply equally well to SEIR and other models that differ from the SIR model only in their distributions of generation intervals (section~\ref{GI}).

Before considering specific cases, we first derive general expressions for the reproduction number and final size of a heterogeneous epidemic. Combining equations~(\ref{eqn:s_eqn})~and~(\ref{eqn:r_eqn} and eliminating the time variable, the system can be described by the set of equations
\begin{equation}
\label{eq:dsdr}
d s_k=-s_k\sum_l R_{kl}dr_l
\end{equation} 
where $R_{kl}\equiv B_{kl}/\gamma_l$.  The basic reproduction number for such a system is given by the top eigenvalue of the next-generation matrix $G$~\cite{Diekmann1990, Driessche2002}; at the beginning of the epidemic, $s_k\approx n_k$, and so
\begin{equation}
\label{eq:g}
    G_{kl}=n_kR_{kl}
\end{equation}
As discussed in the previous section, one can always find parameters $\beta$ and $\gamma$ for a homogeneous SIR model that will reproduce this reproduction number, as well as the initial exponential growth rate of the epidemic.  However, the later trajectory and final size may differ. Specific cases will be given below; a general equation for the final size of a heterogeneous epidemic can be derived as follows.  

Denoting the final size of the epidemic in group $k$ by $r_k^\infty\equiv r_k(t\rightarrow \infty)$ and noting that $s_k(t\rightarrow \infty)=n_k-r_k^\infty$, integrating equation~(\ref{eq:dsdr}) yields
\begin{equation}
\label{eq:int}
\int_{s_k(0)}^{n_k-r_k^\infty}\frac{ds_k}{s_k}=-\sum_lR_{kl}\int_{r_l(0)}^{r_l^\infty}dr_l
\end{equation}
Assuming a small initial number of infections (i.e. $s_k(0)\approx n_k$ and $~r_k(0)\approx0$), equation~(\ref{eq:int}) gives the following implicit expressions for each $r_k^\infty$:
\begin{equation}
\label{eq:infty}
    \frac{r_k^\infty}{n_k} =  1 - \exp{-\sum_l R_{kl} r_l^\infty}
\end{equation} 

As can be seen, the next-generation matrix $G$ and final sizes $r_k^\infty$ depend on the values of $\gamma_k$ only through the ratio $R_{kl}=B_{kl}/\gamma_l$.  To simplify the analysis in what follows, we therefore take $\gamma_k=\gamma$ for all groups $k$, such that $B_{kl}$ is proportional to $R_{kl}$.  This assumption restricts the distribution of generation intervals to that of a homogeneous SIR model, without placing any restrictions on the overall transmission probabilities between infectious and susceptible individuals (which are given by $R_{kl}$).  By holding the distribution of generation intervals constant, we can focus on effects arising from heterogeneity alone.

The initial epidemic growth can be approximated by the set of linear differential equations
\begin{equation}
    \dv{i_k}{t} \approx \sum_{l} M_{kl} i_l
\end{equation}
such that 
\begin{equation}
\label{eq:m}
{M}_{kl} \equiv n_k B_{kl} - \gamma \delta_{kl}=\gamma(G_{kl}-\delta_{kl})
\end{equation} 
where $\delta_{kl}$ is the Kronecker delta.  The initial exponential growth rate in the number of infections $\lambda_0$ will thus be dominated by the top eigenvalue of $M$ and---as in the homogeneous SIR model---is related to the reproduction number $R_0$ (i.e. the top eigenvalue of $G$) by $\lambda_0=\gamma(R_0-1)$.

\subsubsection{Heterogeneity in infectiousness and susceptibility } \label{het_spread}
We first consider homogeneous connectivity (heterogeneous connectivity will be considered in the next subsection) but potentially heterogeneous susceptibility and infectiousness, i.e. we consider $B_{kl}$ that can be factored as $B_{kl}=\eta_k\beta_l$.  We analyze three cases.  

In the first case, groups are equally susceptible but differ in infectiousness, i.e. $B_{kl}=\beta_l$.  In this case, we note that $ds_k/ds_l=s_k/s_l$ and thus the fraction of susceptible individuals in each group will remain constant over the course of the trajectory.   Furthermore, apart from an initial exponentially decaying transient, the ratio of infectious individuals in each group will equal the ratio of susceptible individuals, as can be seen using the time-independence of $s_l/s_k$ to show that 
\begin{equation}
\frac{d}{dt}(i_k\frac{s_l}{s_k}-i_l)=-\gamma(i_k\frac{s_l}{s_k}-i_l).
\end{equation}
Thus we can write 
\begin{equation}
s_k(t)=\frac{s_k(0)}{s(0)}s(t)~~\text{and}~~i_k(t)=\frac{s_k(0)}{s(0)}i(t),
\end{equation}
where $s(t)=\sum_k s_k(t)$ and $i(t)=\sum_k i_k(t)$ are solutions to an SIR model with spreading rate \begin{equation}
    \beta_{\text{SIR}} = \langle \beta \rangle\equiv\sum_k \frac{s_k(0)}{s(0)}\beta_k. 
\end{equation} 
In other words, heterogeneous infectiousness by itself has no impact on the epidemic trajectory.  More generally, in the limit of a large population in which stochastic effects average out, heterogeneous infectiousness can always be incorporated into an otherwise homogeneous model: if each individual has probability $p_i$ of belonging to a group $i$ (such that $\sum_ip_i=1$), with the infectiousness as a function of time of an individual in group $i$ who was infected at time $t=0$ given by $R_0^ig_i(t)$ (such that $\int_0^\infty g_i(t)dt=1$), then the result will be equivalent to a completely homogeneous model with basic reproduction number $R_0=\sum_ip_iR_0^i$ and distribution of generation intervals 
\begin{equation}
    g(t)=\frac{1}{R_0}\sum_ip_iR_0^ig_i(t).
\end{equation}

For instance, the SEIAR model adds an asymptomatic compartment to the SEIR model (e.g. ref.~\cite{Li2021}).   In such a model, asymptomatic individuals cause infections at a different rate than symptomatic individuals, but everyone is equally susceptible and connectivity is homogeneous.  Thus, the SEIAR model by itself has no impact on the epidemic trajectory---an identical trajectory could be obtained by considering an SEIR model and modifying its parameters to account for asymptomatic infection, contact tracing, quarantine/isolation efforts, etc. 

In the second case, the groups can have the same infectiousness but differing susceptibilities, i.e. $B_{kl}=\eta_k\beta$.  By selecting the effective spreading rate $\beta_{\text{SIR}} = \beta \langle \eta \rangle$, the initial growth rate and basic reproduction number can be reproduced with an SIR model (see equation~(\ref{eq:m})). However, the later parts of the epidemic trajectories will diverge, and the total final size $r_\infty=\sum_k r_k^\infty$ is smaller than what would be predicted from a homogeneous model:
\begin{equation}
r_\infty=1-\langle \exp[-(\beta/\gamma)\eta_k r_\infty]\rangle<1-\exp[-(\beta/\gamma)\langle \eta_k\rangle r_\infty] 
\end{equation}
(where the equality follows from equation~(\ref{eq:infty}) and the inequality follows from the concavity of the function $1-e^{-x}$).  

In the third case, both infectiousness and susceptibility vary across groups. We consider a subset of this scenario in which infectiousness and susceptibility are proportional, i.e. those who are more likely to spread the disease are also more likely to contract it. For instance, a person who wears a mask more often or who socializes less will be both less likely to spread and less likely to contract the disease. Assuming both susceptibility and infectiousness are proportional to a contact parameter $b$, i.e. $B_{kl}=b_kb_l$, the homogeneous SIR model can reproduce the initial growth rate and basic reproduction number by selecting the effective spreading rate $\beta_{\text{SIR}} =  \langle b^2 \rangle$ (equation~(\ref{eq:m})). (Note that here, the effective spreading rate $\langle b^2 \rangle$ differs from the average spreading rate $\langle b \rangle^2$ due to the more infectious individuals being more likely to be infected.) However, as in the previous case, the homogeneous SIR model can grossly misestimate later parts of the trajectory and the final epidemic size; theorem 4 of ref.~\cite{Andreasen2011} proves that for a given $R_0$, when susceptibility and infectiousness are proportional to each other, the final size is less than or equal to that of a homogeneous epidemic.  

\begin{figure}%[h]
    \centering
    \includegraphics[scale=1, trim={0 0 0 0}, clip]{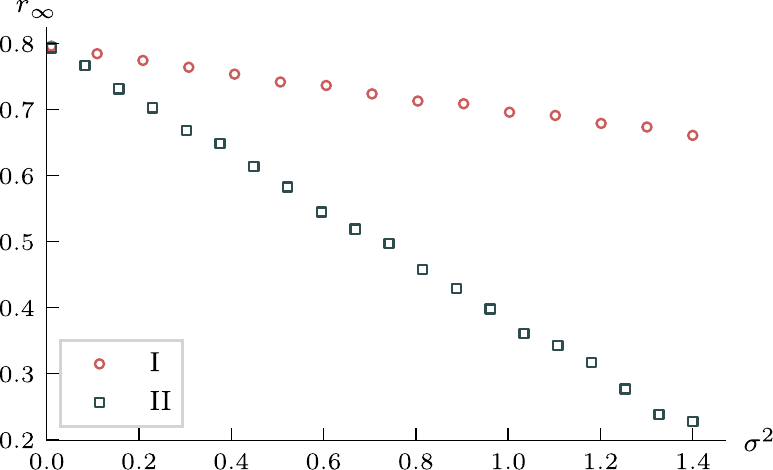}
    \caption{(I) Effect of heterogeneous susceptibility on the final size. Infectiousness is kept constant, while the susceptibility is sampled from gamma distributions with differing variance $\sigma^2$ but with identical means (corresponding to identical values of $R_0$). 
    (II) Effect of heterogeneous susceptibility and infectiousness on the final size, when infectiousness is proportional to susceptibility. The contact parameter to which they are both proportional is sampled from gamma distributions with differing variance $\sigma^2$ but with the same second moment (so as to keep $R_0$ constant).
    In both cases, $R_0=2$, the population consists of 500 equally sized groups, and equation~(\ref{eq:infty}) was used to calculate the final epidemic size $r_\infty=\sum_kr_k^\infty$.
    }
    \label{fig:het_2_types}
\end{figure}

Figure \ref{fig:het_2_types} summarizes these results and shows that the final size can be very different despite identical initial exponential growth rates and basic reproduction numbers (the case of only heterogeneous infectiousness is not shown, as its results are identical to those of an SIR model). Thus, unless individuals are equally likely to be infected, the large-scale effects of heterogeneity on the epidemic trajectory beyond the initial exponential growth cannot be captured by a homogeneous model.  In other words, such heterogeneity cannot be coarse-grained into a single effective spreading rate.

\subsubsection{Heterogeneous connectivity} \label{het_connection}
\begin{figure}%[h]
    \centering
    \includegraphics[scale=1, trim={0 0 0 0}, clip]{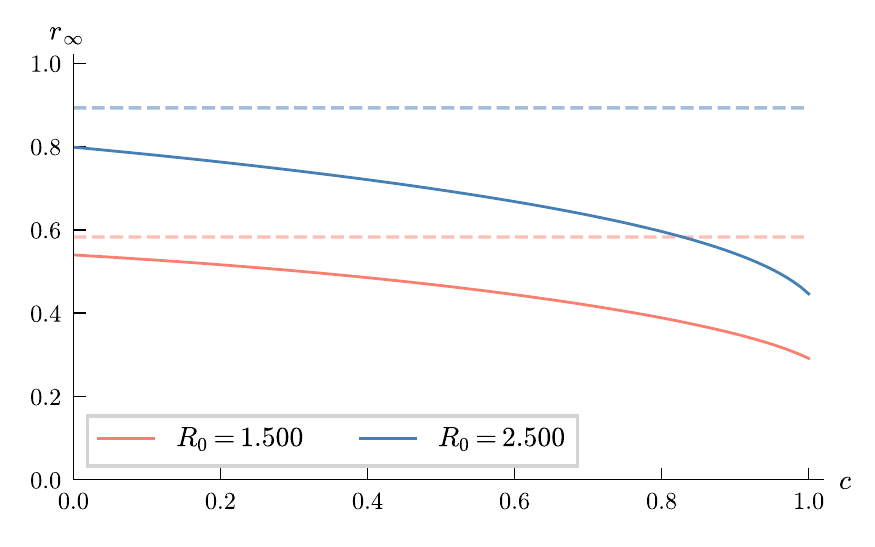}
    \caption{
    Effect of heterogeneous connectivity, infectiousness, and susceptibility on the final epidemic size. For a given value of $R_0$, varying the clustering parameter $c$ from 0 to 1 (and adjusting the contact parameter $b_1$ so as to maintain the same value of $R_0$) in a population containing two groups can lead to epidemics of different sizes. The dashed lines show the epidemic sizes for the same values of $R_0$ in a completely homogeneous model.
    Parameter values are $b_2 = 0.9, n_1 =n_2= 0.5, \gamma = 1.0$ (see equation~\ref{eq:twosubgroups}); equation~(\ref{eq:infty}) was used to calculate the final epidemic size $r_\infty=\sum_kr_k^\infty$.
    }
    \label{fig:c_2}
\end{figure}

The SIR model assumes mean-field connectivity (i.e. every individual is equally likely to interact with every other individual).  Above, we systematically considered heterogeneity in individual characteristics; here, we consider one example of heterogeneous connectivity in which the connectivity within and between groups can be controlled through a clustering parameter between zero and one.  A clustering parameter of zero means that the groups are perfectly well-mixed while a clustering parameter of one means that there is no inter-group disease transmission (mathematical methods can be found in Appendix~\ref{sec:hetcon}).  %A minimal model consisting of two subgroups that also differ in their contact parameters is sufficient to demonstrate that connectivity causes a nonlinear change in the epidemic size. 
Figure~\ref{fig:c_2} provides one example of how connectivity assumptions can affect epidemic size.

More important, however, is the space of policy responses that is opened up by the fact that connectivity is not mean-field (i.e. that populations are not well-mixed).  The geographic clustering of cases, which can be increased with travel restrictions, can be especially helpful in containing a pandemic using only local, targeted measures~\cite{Siegenfeld2020_Commphys}.

\begin{figure}%[h]
    \centering
    \includegraphics[scale=1, trim={0 0 0 0}, clip]{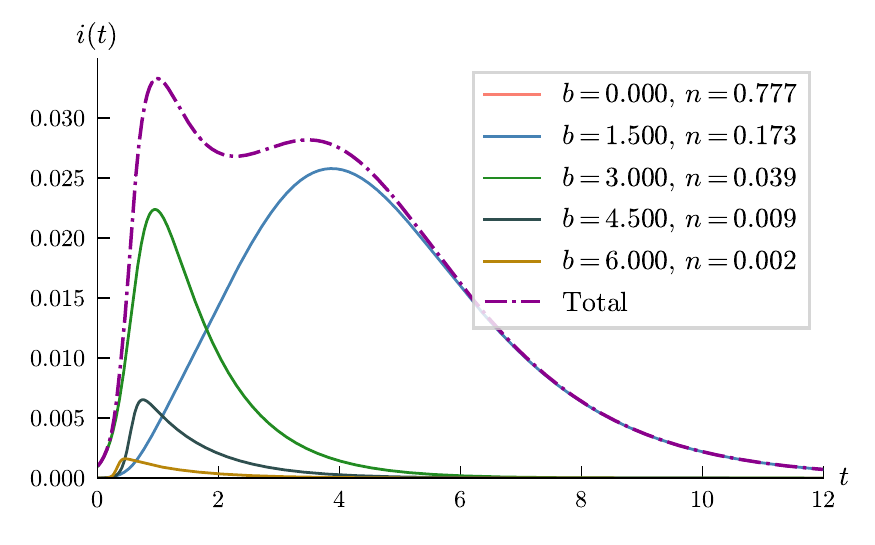}
    \caption{Heterogeneous connectivity, susceptibility, and infectiousness can substantially change the trajectory of the epidemic. Since the groups are well separated, each group exhibits a unique growth rate. If a homogeneous compartmental model was used to forecast the trajectory at $t \sim 1.5$, we would be led to believe that the epidemic was about to end.
    Parameters: $c = 0.75$, $\gamma = 1$, number of groups = $5$, contact parameters $b$ are approximately exponentially distributed with mean $1$ (see Appendix \ref{sec:hetcon}). The seed infection is in the group with $b = 3$.}
    \label{fig:connectivity}
\end{figure}

Another large-scale effect of heterogeneity is that the epidemic trajectory can have multiple peaks and plateaus, an impossible occurrence under homogeneous compartmental models (see Figure \ref{fig:connectivity}). Of course, the shape of an epidemic trajectory will also be affected by policy interventions, behavioral changes in the population, evolution of the infectious agent, and seasonal effects, as well as nonlinear interactions among and between these factors and the various types of heterogeneity.     In Figures \ref{app:fig:plateau_cg}~and~\ref{app:fig:plateau_sweden} (see Appendix), we present time-series of COVID-19 cases in India and Sweden that show large deviations from what a homogeneous model would predict. 

\subsection{Stochasticity and the elimination of outbreaks} \label{stochastic}
\begin{figure}%[h!]
    \centering
    \includegraphics[scale=1.0, trim={0 0 0 0}, clip]{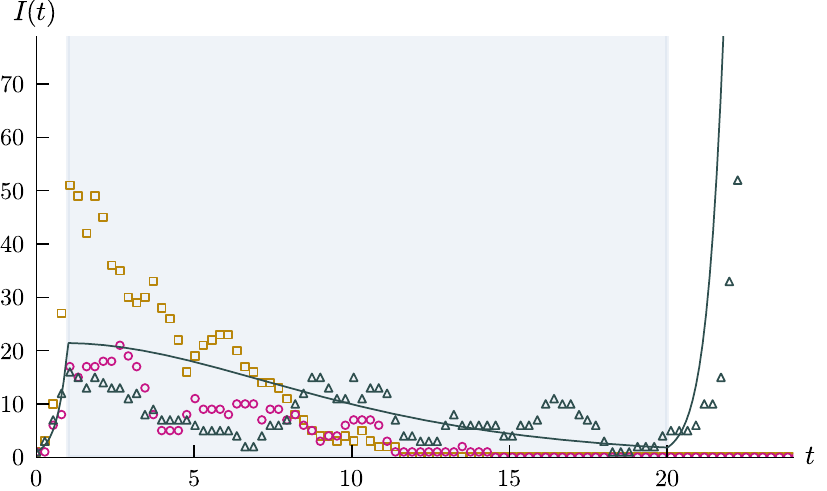}
    \caption{Continuous compartmental models forecast a deterministic second wave of infections. The shaded region of the plot shows the time period for which the spreading rate is reduced as a result of policy interventions. The blue curve shows the number of infections according to a continuous SIR model: after the interventions are removed, the infections rise again. The scatter plot trajectories show the number of infections in a stochastic SIR model, with each marker type corresponding to a single realization of the model. The trajectories marked by pink circles and yellow squares show that elimination is possible and that a second wave need not occur.  Due to the stochastic nature of disease spread, interventions cannot be held in place for a pre-determined amount of time but rather must be calibrated to real-time observations.  For instance, for the cases of the pink circles and yellow squares, the interventions could be lifted earlier than they were in this simulation, while for the blue triangles, the interventions would need to be kept in place for longer.  Simulation details and parameters can be found in Appendix~\ref{app:sto}.}
    \label{app:fig:stochastic}
\end{figure}

In compartmental models, which use continuous variables, the number of infections can exponentially decay but will never reach zero. Such models may mischaracterize the effects of temporary, strong interventions by predicting an inevitable  ``second wave'' \cite{Walker2020, Davies2020, Ferguson2020}. Stochastic compartmental models~\cite{Allen2008} are better suited for analyzing such interventions since the dynamics of disease transmission cannot be approximated as continuous when the number of cases is small.  One behavior that is not captured by most continuous models is the possibility of elimination, i.e. of the infectious fraction of the population equalling zero. Zero is a special number here since any non-zero fraction---no matter how small---can exponentially grow  (see Figure~\ref{app:fig:stochastic} for details).    %Post elimination, a new epidemic can occur only if new cases are imported through travel: a scenario that continuous models do not take into account and forecast a deterministic second wave of infections (if the interventions were imposed before reaching herd immunity threshold). 

Stochastic models also show that not all outbreaks grow to become an epidemic \cite{Althouse2020}, an observation which can aid in identifying policies that achieve containment once cases have been brought to a sufficiently low number. The effect of stochasticity can be particularly pronounced if super-spreader events play a substantial role in the overall spread of the disease. 

Since stochastic disease transmission events take place through the contact networks of individuals, connectivity patterns can affect the dynamics of elimination. Decreasing the effect of long-range connections (e.g. through travel restrictions, testing at borders, etc.) can lead to localized epidemics that are largely independent of one another. These local epidemics are smaller and thus more subject to stochastic effects.  %A homogeneous model would mischaracterize such a state by assuming that the disease is uniformly prevalent across the population, while in reality, some geographical regions would have reached elimination. 
Thus, heterogeneous connectivity can interact with stochasticity to make elimination a more accessible prospect than homogeneous models would imply.

\section{Discussion}
What differentiates a good model from a bad model is not its level of detail but rather the relationship between the details included in the model and the most important behaviors of the system.  Which details are important can depend not only on the system but also on the modelling objectives.  For instance, the distribution of generation intervals is crucial if attempting to calculate the reproduction number from the exponential growth/decline rate of an epidemic, but it can generally be coarse-grained to a single time-scale in the context of predicting overall epidemic trajectories. 

%SEIR models differ from the SIR model in that they use a more detailed and realistic generation interval distribution, the corrections to the epidemic trajectory from which are generally swamped by other sources of error.  
SEIR and many other compartmental models\cite{TangLu2020, roosa2019, mehdaoui2021, brauer2008} differ from the SIR model in that they use a more detailed and realistic generation interval distribution, and in some cases have heterogeneous infectiousness (as in SEIAR models).  However, as shown in sections~\ref{GI}~and~\ref{het}, the corrections to the epidemic trajectory from the distribution of generation intervals will generally be small compared to other sources of error, and heterogeneity in infectiousness alone has no impact in these deterministic models.  Both SIR and SEIR models ignore potentially important factors such as heterogeneous connectivity, stochasticity, and behavior change/policy interventions.  If such factors are to be ignored, however, the SIR model has the advantage of not including any unnecessary (and therefore potentially misleading) details; its output depends only on the unitless $R_0$, together with a time-scale set by the effective recovery rate $\gamma$.
%While the details that are present in an SEIR class model do not affect the large scale behaviors, it does not mean that the breakdown of types of infections (quarantined, asymptomatic, severe or hospitalized) is not useful. Such a breakdown might help with the management of health care resources and with analyzing interventions targeted at individuals in specific disease stages.
SEIR and related models can provide a breakdown of types of infections and can help with, for example, the management of health care resources (although often, all that matters is the probability of an infection being of a certain type rather than the precise dynamics between types). But given the often far larger effects of heterogeneity and stochasticity (not to mention behavioral change and policy response), including the details of disease progression while ignoring these other assumptions may provide a false sense of confidence in the accuracy of the model.  %Furthermore, the use of stochastic, discrete variables in place of the continuous variables used in compartmental models, allows for behaviors such as elimination in finite time (e.g. due to social distancing) that cannot otherwise be captured. 
More significantly, a misunderstanding of the relative importance of assumptions in any given model may narrow the set of interventions considered. For instance, the effect of travel restrictions cannot even be described in a model with homogeneous connectivity.  %Both stochasticity and heterogeneity in connectivity (especially that which is geographic) greatly open up the space of possible interventions.  

The idea that some details and assumptions are more important than others is frequently used in mathematics and physics. Functions are often approximated using a power-series expansion, with each higher-order term providing additional details. A higher-order term (finer-grain correction) is used only when all lower-order terms (coarser-grain corrections) have already been included.  To do otherwise---or to include some corrections at a given scale while ignoring others---is mathematically unsound and can lead to nonsensical results.  When modeling many real-world systems, the various details do not necessarily fit cleanly into a power series, but the general conceptual principle still holds: details of lesser relative importance should be considered only after all of the larger-scale effects have already been taken into account. 

Agent-based or network models can transcend some of the limitations of compartmental models.  Like any model, however, they may suffer from the flaw of arbitrarily focusing on some details while leaving out others, thereby mischaracterizing the space of large-scale behaviors of the epidemic.   As agent-based and network models are generally more detailed, especially careful attention must be paid to this point. 
%For instance, agent based modeling may include highly detailed contact networks, travel networks, details of disease progression etc. But the descriptions of these details will necessarily be imperfect and other equally important details may not be included.  Thus, more important than an agent-based or network model capturing all of these details is for it to characterize space of possible policy responses (e.g. social distancing measures, travel restrictions) and their effect on the large-scale behavior of an epidemic (e.g. whether or not the epidemic is eliminated). 

In addition to the assumptions that we have analyzed in detail, a few general comments can be made about the potentially large-scale effects of policy response, temporary immunity, and mutation of the pathogen. Policy responses such as social distancing, mask mandates, contact tracing, travel restrictions, mass testing, ventilation, and lock-downs can greatly affect disease transmission and change whether the effective reproduction number is greater than or less than one.  If immunity wanes over time, then recovered individuals can become susceptible again (as captured using SIRS and related models), such that, in the absence of elimination efforts, the disease will become endemic and will not have a final size.  The mutation of a pathogen can render the immunity developed to previous strains less effective and can also substantially change the transmission characteristics of the disease. 
And there are many other details that we have not examined such as contraction of generation intervals \cite{Kenah2008}, vital dynamics, and seasonal effects, among others, which may influence the  behaviors of an epidemic in various ways.  This unpredictability inherent to epidemics underscores the need for a precautionary principle for acting under uncertainty \cite{Cirillo2020}.  A careful examination of model assumptions is necessary, not to evaluate the accuracy of the assumptions themselves---they will always be inaccurate---but to see how they do or do not affect the link between our actions and the space of possible outcomes.  %Accurately characterizing heterogeneity is impossible
%\footnote{Mobility patterns can be obtained, but fine scale details like how often each individual wears a mask, distribution of viral load, spatial distance between individuals etc. are nearly impossible to obtain. Our model and other models that incorporate heterogeneity coarse grain over these details through effective parameters, and whether these parameters can capture the large scale effects is not known.}

%Models can illuminate the space of possible outcomes and interventions and provide guidance as to how these outcomes are linked to our actions.

\section{Conclusion}
Our analysis indicates that compartmental models often focus on details that do not matter while ignoring far larger sources of error.  More precisely, when compartmental models do include additional details, they tend to focus on the distribution of generation intervals and sometimes heterogeneous infectiousness, while ignoring other assumptions that have far larger effects.  Including corrections in a model that has gotten the big picture wrong, however, serves only to give a false sense of confidence.  For instance, stochastic effects can sometimes mean the difference between the existence or non-existence of a disease within a population.  Heterogeneity in connectivity can open up entirely new regimes of policy control, including internal travel restrictions, and, even in the absence of such policy, such heterogeneity, especially when combined with other sources of heterogeneity, can dramatically alter an epidemic's trajectory and size.  In comparison, without also including stochasticity, heterogeneity in infectiousness alone has no impact, and the addition of disease stages will only affect the precise timing of disease spread.  This timing can be important, but can also be approximately captured using the simpler SIR model with the appropriate choice of recovery rate, which together with the other transition rates between compartments should be thought of as effective parameters rather than as actually modeling an exponential distribution of time spent in each compartment.  None of this is to say that models with additional compartments should not be used; rather, careful attention must be paid to whether each additional parameter can actually better guide action, given the context of both the other assumptions of the model and the sources of uncertainty within the data.

\subsection*{Acknowledgements}
\vspace{-1.35em}
This material is based upon work supported by the National Science Foundation Graduate Research Fellowship Program under Grant No. 1122374 and by the Hertz Foundation. 
\vspace{-1.5em}
%\subsection*{Disclosure}
%\vspace{-1.35em}
%A preprint has been previously published~\cite{preprint}.
%\clearpage
%\onecolumngrid
\appendix
\section*{Appendix} 

\subsection{Epidemic size} \label{app:size}
The final size of an epidemic in which each individual can be infected at most once is not affected by the distribution of generation intervals. Consider an epidemic process in a population of $N$ individuals, where each infected individual has a probability $p$ of infecting any given susceptible individual and where these infection events are independent of each other. Regardless of the distribution of generation intervals, the transmission events of this process can be represented by an Erdos-Renyi graph \cite{barbour1990, ball1997}. The size of the epidemic $r_\infty$ will then be the fraction of nodes  within the giant component of the Erdos-Renyi graph, with an epidemic occurring whenever (one of) the seed infection(s) is part of this giant component.   Thus, under assumptions of homogeneity, the epidemic size depends only on $p$ and $N$.  In the limit as $N\rightarrow\infty$ with $R_0 = pN$ held constant, the fraction of nodes in the giant component is given by $r_\infty = 1 - \exp(-R_0 r_\infty)$, which, as expected, is the same result given by deterministic homogeneous compartmental models. 

Heterogeneity can be incorporated by considering a set of nodes on which each transmission event is represented by a directed edge. The transmission events need not be independent. The probability that any given individual will be infected is simply equal to the probability that that individual is connected directly or indirectly to a seed infection node.  Thus, the final size (and, indeed, each individual's probability of at some point being infected) depends only on the probability of transmission between each ordered pair of individuals conditioning on the first  being infectious and the other susceptible---i.e. on the probability of a directed edge existing between the pair of nodes---as well as the seed infections, and does not depend on the temporal properties of the process such as generation intervals or the number or types of stages in a compartmental model~\cite{Newman2002}.

\subsection{The SEIR model} \label{app:SEIR}
The SEIR model is a modified SIR model in which a new compartment of exposed individuals (who have been infected but are not yet infectious) is introduced.  The model is described by these differential equations:
\begin{align}
    \dv{s}{t} &= -\beta s i \\
    \dv{e}{t} &= \beta s i - \sigma e \label{eq:e}\\
    \dv{i}{t} &= \sigma e - \alpha i \label{eq:i}\\
    \dv{r}{t} &= \alpha i
\end{align}
Convolving the two exponential distributions corresponding to the transitions from exposed to infectious and infectious to recovered~\cite{Wallinga2006} gives the following distribution of generation intervals
\begin{align}
    g(t) = \frac{\sigma \alpha}{\sigma - \alpha} (e^{-\alpha t} - e^{-\sigma t})
\end{align}
which, when combined with equation~(\ref{eqn:Wallinga}), yields
\begin{equation}
\label{eqn:SEIR}
    R_0 = (1+\lambda_0/\sigma)(1+\lambda_0/\alpha)
\end{equation}
where $\lambda_0$ is the initial exponential growth rate and $R_0$ is the basic reproduction number.  Combining equations~(\ref{eqn:SEIR}) and~(\ref{eqn:Wallinga_SIR}) allows us to find an effective $\gamma$ of the SIR model in terms of the effective parameters of the SEIR model such that the basic reproduction number and the initial growth rate of both the models are equal:
\begin{align}
    R_0 &= (1+\lambda_0/\sigma)(1+\lambda_0/\alpha) = 1+\lambda_0/\gamma %\\
    %\implies \gamma_{\text{eff}} 
    %&=\bigg ( \frac{1}{\lambda} \bigg ( 1+\frac{\lambda}{\sigma}\bigg ) \bigg( 1+\frac{\lambda}{\alpha}\bigg) - 1\bigg )^{-1}
\end{align}
Thus, any SEIR model can be replaced with an SIR model with the same initial growth rate and $R_0$ (and thus final size).  (Note that for both models, $\beta$ is also an effective parameter: for the SEIR model, $\beta=R_0\alpha$, while for the SIR model, $\beta=R_0\gamma$.) The two models will differ only in terms of the precise timing the epidemic curve later on in its trajectory, but such differences will be swamped by other sources of error such as heterogeneity and stochasticity.  

\begin{figure}%[h!]
    \centering
    \includegraphics[scale=.9, trim={0 0 0 0}, clip]{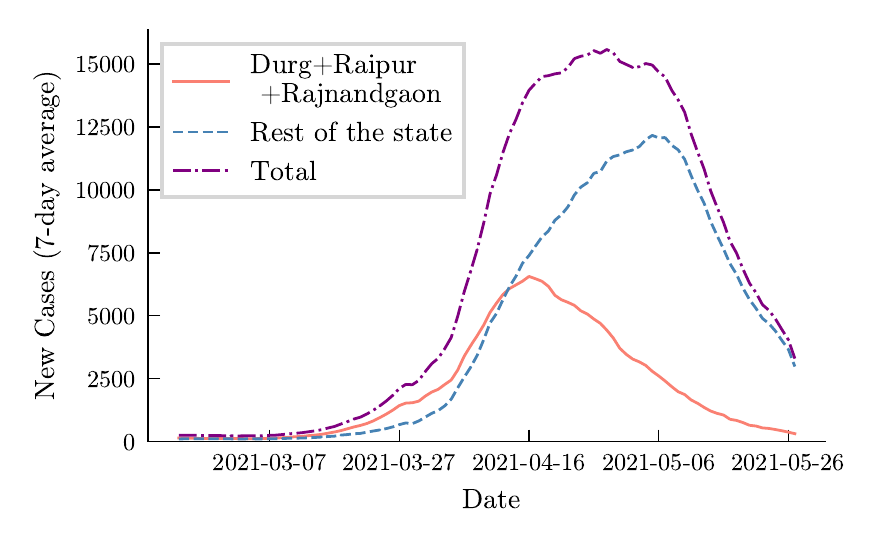}
    \vspace{-1em}
    \caption{Time series of daily new COVID-19 infections from the state of Chhattisgarh in India shows an epidemic plateau between 2021-04-16 and 2021-05-06. Three districts of the state (out of 27)---Durg, Raipur and Rajnandgaon---peaked around 2021-04-16, while the rest of the districts in the state peaked around 2021-05-06. This disparity in the trajectories is a result of the heterogeneity in the social contact structures, as well as in the timing and strength of policy and behavioral changes.  Epidemic data was obtained from ref.~\cite{covid19india}.}
    \label{app:fig:plateau_cg}
\end{figure}

\begin{figure}%[h!]
    \centering
    \includegraphics[scale=.9, trim={0 0 0 0}, clip]{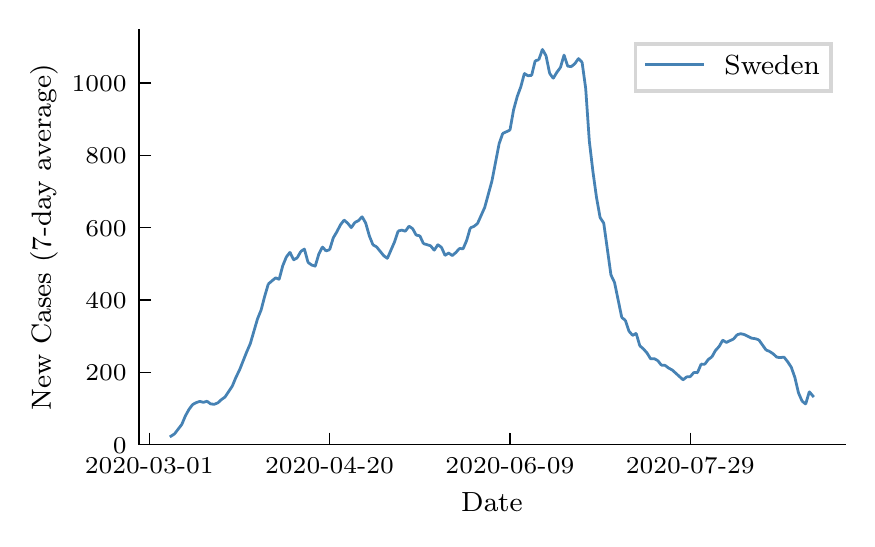}
    \vspace{-1em}
    \caption{Time series of daily new COVID-19 infections from Sweden shows an epidemic plateau in April and May 2020. In the absence of sharp policy/behavioral changes, a homogeneous compartmental model cannot account for such an early plateau, which was clearly not the result of population-wide herd immunity.  In a heterogeneous model, such a plateau could be partially explained by substantial immunity within the specific sub-populations that were responsible for the initial exponential growth.   Epidemic data was obtained from refs.~\cite{owidcoronavirus, dong2020interactive}}
    \label{app:fig:plateau_sweden}
\end{figure}

\subsection{Heterogeneous connectivity} \label{sec:hetcon}
Under the assumption of homogeneous connectivity, the rate $B_{kl}$  at which an individual in group $k$ infects an individual in group $l$ (see equations~(\ref{eqn:s_eqn})-(\ref{eqn:r_eqn})) can be described solely in terms of the individual characteristics of members of groups $k$ and $l$.  But in reality, transmission tends to be clustered.   In order to capture some of these clustering effects (albeit in a simplified manner), we allow for a higher probability of the disease spreading within groups than between groups.  For instance, it is generally more likely for the disease to spread between two individuals living within the same city than between individuals living in different cities. 

To account for this, we let $B_{kl}=b_kb_lC_{kl}$ with $C_{kl}$ given by
\begin{align} \label{eqns_basic}
   C_{kl} =  1-c + \frac{c}{n_k}\delta_{kl}
\end{align} 
where $c\in [0,1]$ is the clustering parameter.  Note that $C_{kl}$ is normalized such that for all $l$, $\sum_{k} n_k C_{kl} = 1$.

In Figure \ref{fig:c_2} of the main text, we explore the case of two groups with different contact parameters $b_1$ and $b_2$ and a connectivity parameter $c$. The basic reproduction number $R_0$ is then given by 
\begin{align}
\label{eq:twosubgroups}
    R_0 &= \frac{1}{2\gamma}(n_1 B_1 + n_2 B_2 + \sqrt{(n_1 B_1 - n_2 B_2)^2 + 4n_1 n_2 B_c^2})
\end{align} 
where  $B_1 = b_1^2(1-c+c/n_1)$,~~~$B_2 = b_2^2(1-c+c/n_2)$, and $B_c = b_1 b_2 (1-c)$.

Heterogeneous connectivity can result in deviation from simple epidemic trajectories of  growth followed by decline, as shown in Figure \ref{fig:connectivity} of the main text. One such deviation is a plateau in the epidemic trajectory observed in both simulations \cite{Maltsev2021} and epidemic data. Figure \ref{app:fig:plateau_cg} shows such a plateau for a wave of COVID-19 in Chhattisgarh, India. Figure \ref{app:fig:plateau_sweden} shows COVID-19 cases in Sweden, in which the number of infections did not decline after the curve flattened, despite the absence of major changes in the mitigation policies.

\vspace{.5em}
\subsection{Stochasticity} \label{app:sto}
\vspace{-.5em}
For Figure~\ref{app:fig:stochastic}, we use the discrete-time Markov chain (DTMC) formulation for simulating a stochastic epidemic \cite{Allen2008}. A discrete population of size $N$ is described by the state $(S, I)$, where $S$ is the number of susceptible individuals and $I$ is the number of infected individuals. Similar to the deterministic SIR model, $\beta$ and $\gamma$ are the effective spreading and recovery rates. The simulation starts with a single infected individual---i.e. the population starts in the state $(N-1, 1)$---with transition probabilities from the state $(S, I)$ given by 
\begin{align}
    P(S-1, I+1) &= \beta \frac{S I}{N} \Delta t \\
    P(S, I-1) &= \gamma I \Delta t \\
    P(S, I) &= 1 - \Big (\beta \frac{S}{N} + \gamma \Big) I \Delta t
\end{align}
where $\Delta t$ is the time step for the simulation. Its value must be selected such that all transition probabilities lie between zero and one.  In Figure~\ref{app:fig:stochastic}, $\beta = 4.1, \gamma = 1, N = 1000$, and $\Delta t = 0.00017$, with $\beta$ being reduced by a factor of 4 during the intervention period, which lasts from $t = 1.0$ to $t = 20.0$. 

%Figure \ref{app:fig:stochastic} shows three instances of this simulation

%\clearpage
\bibliography{references}{}

\begin{thebibliography}{10}
\expandafter\ifx\csname url\endcsname\relax
  \def\url#1{\texttt{#1}}\fi
\expandafter\ifx\csname urlprefix\endcsname\relax\def\urlprefix{URL }\fi
\providecommand{\bibinfo}[2]{#2}
\providecommand{\eprint}[2][]{\url{#2}}

\bibitem{Hethcote2000}
\bibinfo{author}{Hethcote, H.~W.}
\newblock \bibinfo{title}{The mathematics of infectious diseases}.
\newblock \emph{\bibinfo{journal}{SIAM {R}eview}}
  \textbf{\bibinfo{volume}{42}}, \bibinfo{pages}{599--653}
  (\bibinfo{year}{2000}).

\bibitem{Rock2014}
\bibinfo{author}{Rock, K.}, \bibinfo{author}{Brand, S.}, \bibinfo{author}{Moir,
  J.} \& \bibinfo{author}{Keeling, M.~J.}
\newblock \bibinfo{title}{Dynamics of infectious diseases}.
\newblock \emph{\bibinfo{journal}{Reports on Progress in Physics}}
  \textbf{\bibinfo{volume}{77}}, \bibinfo{pages}{026602}
  (\bibinfo{year}{2014}).
\newblock \urlprefix\url{https://doi.org/10.1088/0034-4885/77/2/026602}.

\bibitem{Barrat2008}
\bibinfo{author}{Barrat, A.}, \bibinfo{author}{Barthelemy, M.} \&
  \bibinfo{author}{Vespignani, A.}
\newblock \emph{\bibinfo{title}{Dynamical Processes on Complex Networks}}
  (\bibinfo{publisher}{Cambridge University Press}, \bibinfo{year}{2008}).
\newblock \urlprefix\url{https://doi.org/10.1017/cbo9780511791383}.

\bibitem{PastorSatorras2015}
\bibinfo{author}{Pastor-Satorras, R.}, \bibinfo{author}{Castellano, C.},
  \bibinfo{author}{Mieghem, P.~V.} \& \bibinfo{author}{Vespignani, A.}
\newblock \bibinfo{title}{Epidemic processes in complex networks}.
\newblock \emph{\bibinfo{journal}{Reviews of Modern Physics}}
  \textbf{\bibinfo{volume}{87}}, \bibinfo{pages}{925--979}
  (\bibinfo{year}{2015}).
\newblock \urlprefix\url{https://doi.org/10.1103/revmodphys.87.925}.

\bibitem{Billah_2020}
\bibinfo{author}{Billah, M.~A.}, \bibinfo{author}{Miah, M.~M.} \&
  \bibinfo{author}{Khan, M.~N.}
\newblock \bibinfo{title}{Reproductive number of coronavirus: A systematic
  review and meta-analysis based on global level evidence}.
\newblock \emph{\bibinfo{journal}{PLOS ONE}} \textbf{\bibinfo{volume}{15}},
  \bibinfo{pages}{e0242128} (\bibinfo{year}{2020}).
\newblock
  \urlprefix\url{https://journals.plos.org/plosone/article?id=10.1371/journal.pone.0242128}.

\bibitem{Tolles2020}
\bibinfo{author}{Tolles, J.} \& \bibinfo{author}{Luong, T.}
\newblock \bibinfo{title}{Modeling epidemics with compartmental models}.
\newblock \emph{\bibinfo{journal}{{JAMA}}} \textbf{\bibinfo{volume}{323}},
  \bibinfo{pages}{2515} (\bibinfo{year}{2020}).
\newblock \urlprefix\url{https://doi.org/10.1001/jama.2020.8420}.

\bibitem{Wolfram_SIR}
\bibinfo{author}{Weisstein, E.~W.}
\newblock \bibinfo{title}{Kermack-{M}ckendrick model}.
\newblock
  \bibinfo{howpublished}{\url{https://mathworld.wolfram.com/Kermack-McKendrickModel.html}}.
\newblock \bibinfo{note}{Accessed: 2021-04-17}.

\bibitem{Roberts2015}
\bibinfo{author}{Roberts, M.}, \bibinfo{author}{Andreasen, V.},
  \bibinfo{author}{Lloyd, A.} \& \bibinfo{author}{Pellis, L.}
\newblock \bibinfo{title}{Nine challenges for deterministic epidemic models}.
\newblock \emph{\bibinfo{journal}{Epidemics}} \textbf{\bibinfo{volume}{10}},
  \bibinfo{pages}{49--53} (\bibinfo{year}{2015}).
\newblock \urlprefix\url{https://doi.org/10.1016/j.epidem.2014.09.006}.

\bibitem{Givan2011}
\bibinfo{author}{Givan, O.}, \bibinfo{author}{Schwartz, N.},
  \bibinfo{author}{Cygelberg, A.} \& \bibinfo{author}{Stone, L.}
\newblock \bibinfo{title}{Predicting epidemic thresholds on complex networks:
  Limitations of mean-field approaches}.
\newblock \emph{\bibinfo{journal}{Journal of Theoretical Biology}}
  \textbf{\bibinfo{volume}{288}}, \bibinfo{pages}{21--28}
  (\bibinfo{year}{2011}).
\newblock
  \urlprefix\url{https://www.sciencedirect.com/science/article/pii/S0022519311003651}.

\bibitem{DHAR2020}
\bibinfo{author}{Dhar, A.}
\newblock \bibinfo{title}{What one can learn from the {SIR} model}.
\newblock \emph{\bibinfo{journal}{Indian Academy of Sciences Conference
  Series}} \textbf{\bibinfo{volume}{3}} (\bibinfo{year}{2020}).
\newblock \urlprefix\url{https://doi.org/10.29195/iascs.03.01.0016}.

\bibitem{Siegenfeld2020}
\bibinfo{author}{Siegenfeld, A.~F.} \& \bibinfo{author}{Bar-Yam, Y.}
\newblock \bibinfo{title}{An introduction to complex systems science and its
  applications}.
\newblock \emph{\bibinfo{journal}{Complexity}} \textbf{\bibinfo{volume}{2020}},
  \bibinfo{pages}{1--16} (\bibinfo{year}{2020}).
\newblock \urlprefix\url{https://doi.org/10.1155/2020/6105872}.

\bibitem{Siegenfeld2020_PNAS}
\bibinfo{author}{Siegenfeld, A.~F.}, \bibinfo{author}{Taleb, N.~N.} \&
  \bibinfo{author}{Bar-Yam, Y.}
\newblock \bibinfo{title}{Opinion: What models can and cannot tell us about
  {COVID}-19}.
\newblock \emph{\bibinfo{journal}{Proceedings of the National Academy of
  Sciences}} \textbf{\bibinfo{volume}{117}}, \bibinfo{pages}{16092--16095}
  (\bibinfo{year}{2020}).
\newblock \urlprefix\url{https://doi.org/10.1073/pnas.2011542117}.

\bibitem{Siegenfeld2020_Commphys}
\bibinfo{author}{Siegenfeld, A.~F.} \& \bibinfo{author}{Bar-Yam, Y.}
\newblock \bibinfo{title}{The impact of travel and timing in eliminating
  {COVID}-19}.
\newblock \emph{\bibinfo{journal}{Communications Physics}}
  \textbf{\bibinfo{volume}{3}} (\bibinfo{year}{2020}).
\newblock \urlprefix\url{https://doi.org/10.1038/s42005-020-00470-7}.

\bibitem{Miller2012}
\bibinfo{author}{Miller, J.~C.}
\newblock \bibinfo{title}{A note on the derivation of epidemic final sizes}.
\newblock \emph{\bibinfo{journal}{Bulletin of Mathematical Biology}}
  \textbf{\bibinfo{volume}{74}}, \bibinfo{pages}{2125--2141}
  (\bibinfo{year}{2012}).
\newblock \urlprefix\url{https://doi.org/10.1007/s11538-012-9749-6}.

\bibitem{Andreasen2011}
\bibinfo{author}{Andreasen, V.}
\newblock \bibinfo{title}{The final size of an epidemic and its relation to the
  basic reproduction number}.
\newblock \emph{\bibinfo{journal}{Bulletin of Mathematical Biology}}
  \textbf{\bibinfo{volume}{73}}, \bibinfo{pages}{2305--2321}
  (\bibinfo{year}{2011}).
\newblock \urlprefix\url{https://doi.org/10.1007/s11538-010-9623-3}.

\bibitem{Ma2006}
\bibinfo{author}{Ma, J.} \& \bibinfo{author}{Earn, D. J.~D.}
\newblock \bibinfo{title}{Generality of the final size formula for an epidemic
  of a newly invading infectious disease}.
\newblock \emph{\bibinfo{journal}{Bulletin of Mathematical Biology}}
  \textbf{\bibinfo{volume}{68}}, \bibinfo{pages}{679--702}
  (\bibinfo{year}{2006}).
\newblock \urlprefix\url{https://doi.org/10.1007/s11538-005-9047-7}.

\bibitem{ball1997}
\bibinfo{author}{Ball, F.}, \bibinfo{author}{Mollison, D.} \&
  \bibinfo{author}{Scalia-Tomba, G.}
\newblock \bibinfo{title}{Epidemics with two levels of mixing}.
\newblock \emph{\bibinfo{journal}{The Annals of Applied Probability}}
  \textbf{\bibinfo{volume}{7}} (\bibinfo{year}{1997}).
\newblock \urlprefix\url{https://doi.org/10.1214/aoap/1034625252}.

\bibitem{barbour1990}
\bibinfo{author}{Barbour, A.} \& \bibinfo{author}{Mollison, D.}
\newblock \bibinfo{title}{Epidemics and random graphs}.
\newblock In \emph{\bibinfo{booktitle}{Stochastic Processes in Epidemic
  Theory}}, \bibinfo{pages}{86--89} (\bibinfo{publisher}{Springer Berlin
  Heidelberg}, \bibinfo{year}{1990}).
\newblock \urlprefix\url{https://doi.org/10.1007/978-3-662-10067-7_8}.

\bibitem{Wallinga2006}
\bibinfo{author}{Wallinga, J.} \& \bibinfo{author}{Lipsitch, M.}
\newblock \bibinfo{title}{How generation intervals shape the relationship
  between growth rates and reproductive numbers}.
\newblock \emph{\bibinfo{journal}{Proceedings of the Royal Society B:
  Biological Sciences}} \textbf{\bibinfo{volume}{274}},
  \bibinfo{pages}{599--604} (\bibinfo{year}{2006}).
\newblock \urlprefix\url{https://doi.org/10.1098/rspb.2006.3754}.

\bibitem{Choi2020}
\bibinfo{author}{Choi, S.} \& \bibinfo{author}{Ki, M.}
\newblock \bibinfo{title}{Estimating the reproductive number and the outbreak
  size of {COVID}-19 in {K}orea}.
\newblock \emph{\bibinfo{journal}{Epidemiology and Health}}
  \textbf{\bibinfo{volume}{42}}, \bibinfo{pages}{e2020011}
  (\bibinfo{year}{2020}).
\newblock \urlprefix\url{https://doi.org/10.4178/epih.e2020011}.

\bibitem{Hyafil2020}
\bibinfo{author}{Hyafil, A.} \& \bibinfo{author}{Mori{\~{n}}a, D.}
\newblock \bibinfo{title}{Analysis of the impact of lockdown on the
  reproduction number of the {SARS}-{Cov}-2 in {S}pain}.
\newblock \emph{\bibinfo{journal}{Gaceta Sanitaria}}
  \textbf{\bibinfo{volume}{35}}, \bibinfo{pages}{453--458}
  (\bibinfo{year}{2021}).
\newblock \urlprefix\url{https://doi.org/10.1016/j.gaceta.2020.05.003}.

\bibitem{Kuniya2020}
\bibinfo{author}{Kuniya, T.}
\newblock \bibinfo{title}{Prediction of the epidemic peak of coronavirus
  disease in {J}apan, 2020}.
\newblock \emph{\bibinfo{journal}{Journal of Clinical Medicine}}
  \textbf{\bibinfo{volume}{9}}, \bibinfo{pages}{789} (\bibinfo{year}{2020}).
\newblock \urlprefix\url{https://doi.org/10.3390/jcm9030789}.

\bibitem{Read2020}
\bibinfo{author}{Read, J.~M.}, \bibinfo{author}{Bridgen, J.~R.},
  \bibinfo{author}{Cummings, D.~A.}, \bibinfo{author}{Ho, A.} \&
  \bibinfo{author}{Jewell, C.~P.}
\newblock \bibinfo{title}{Novel coronavirus 2019-{nCoV}: early estimation of
  epidemiological parameters and epidemic predictions}  (\bibinfo{year}{2020}).
\newblock \urlprefix\url{https://doi.org/10.1101/2020.01.23.20018549}.

\bibitem{Tang2020}
\bibinfo{author}{Tang, B.} \emph{et~al.}
\newblock \bibinfo{title}{Estimation of the transmission risk of the
  2019-{nCoV} and its implication for public health interventions}.
\newblock \emph{\bibinfo{journal}{Journal of Clinical Medicine}}
  \textbf{\bibinfo{volume}{9}}, \bibinfo{pages}{462} (\bibinfo{year}{2020}).
\newblock \urlprefix\url{https://doi.org/10.3390/jcm9020462}.

\bibitem{Wu2020}
\bibinfo{author}{Wu, J.~T.}, \bibinfo{author}{Leung, K.} \&
  \bibinfo{author}{Leung, G.~M.}
\newblock \bibinfo{title}{Nowcasting and forecasting the potential domestic and
  international spread of the 2019-{nCoV} outbreak originating in {W}uhan,
  {C}hina: a modelling study}.
\newblock \emph{\bibinfo{journal}{The Lancet}} \textbf{\bibinfo{volume}{395}},
  \bibinfo{pages}{689--697} (\bibinfo{year}{2020}).
\newblock \urlprefix\url{https://doi.org/10.1016/s0140-6736(20)30260-9}.

\bibitem{Zhou2020}
\bibinfo{author}{Zhou, H.} \emph{et~al.}
\newblock \bibinfo{title}{Characterizing the transmission and identifying the
  control strategy for {COVID}-19 through epidemiological modeling}.
\newblock \emph{\bibinfo{journal}{medRxiv}}  (\bibinfo{year}{2020}).
\newblock \urlprefix\url{https://doi.org/10.1101/2020.02.24.20026773}.

\bibitem{Kyrychko2020}
\bibinfo{author}{Kyrychko, Y.~N.}, \bibinfo{author}{Blyuss, K.~B.} \&
  \bibinfo{author}{Brovchenko, I.}
\newblock \bibinfo{title}{Mathematical modelling of the dynamics and
  containment of {COVID}-19 in {U}kraine}.
\newblock \emph{\bibinfo{journal}{Scientific Reports}}
  \textbf{\bibinfo{volume}{10}} (\bibinfo{year}{2020}).
\newblock \urlprefix\url{https://doi.org/10.1038/s41598-020-76710-1}.

\bibitem{Walker2020}
\bibinfo{author}{Walker, P. G.~T.} \emph{et~al.}
\newblock \bibinfo{title}{The impact of {COVID}-19 and strategies for
  mitigation and suppression in low- and middle-income countries}.
\newblock \emph{\bibinfo{journal}{Science}} \textbf{\bibinfo{volume}{369}},
  \bibinfo{pages}{413--422} (\bibinfo{year}{2020}).
\newblock \urlprefix\url{https://doi.org/10.1126/science.abc0035}.

\bibitem{Davies2020}
\bibinfo{author}{Davies, N.~G.} \emph{et~al.}
\newblock \bibinfo{title}{Effects of non-pharmaceutical interventions on
  {COVID}-19 cases, deaths, and demand for hospital services in the {UK}: a
  modelling study}.
\newblock \emph{\bibinfo{journal}{The Lancet Public Health}}
  \textbf{\bibinfo{volume}{5}}, \bibinfo{pages}{e375--e385}
  (\bibinfo{year}{2020}).
\newblock \urlprefix\url{https://doi.org/10.1016/s2468-2667(20)30133-x}.

\bibitem{Chowdhury2020}
\bibinfo{author}{Chowdhury, R.} \emph{et~al.}
\newblock \bibinfo{title}{Dynamic interventions to control {COVID}-19 pandemic:
  a multivariate prediction modelling study comparing 16 worldwide countries}.
\newblock \emph{\bibinfo{journal}{European Journal of Epidemiology}}
  \textbf{\bibinfo{volume}{35}}, \bibinfo{pages}{389--399}
  (\bibinfo{year}{2020}).
\newblock \urlprefix\url{https://doi.org/10.1007/s10654-020-00649-w}.

\bibitem{Radulescu2020}
\bibinfo{author}{Rǎdulescu, A.}, \bibinfo{author}{Williams, C.} \&
  \bibinfo{author}{Cavanagh, K.}
\newblock \bibinfo{title}{Management strategies in a {SEIR}-type model of
  {COVID} 19 community spread}.
\newblock \emph{\bibinfo{journal}{Scientific Reports}}
  \textbf{\bibinfo{volume}{10}} (\bibinfo{year}{2020}).
\newblock \urlprefix\url{https://doi.org/10.1038/s41598-020-77628-4}.

\bibitem{Scala2020}
\bibinfo{author}{Scala, A.} \emph{et~al.}
\newblock \bibinfo{title}{Time, space and social interactions: exit mechanisms
  for the {COVID}-19 epidemics}.
\newblock \emph{\bibinfo{journal}{Scientific Reports}}
  \textbf{\bibinfo{volume}{10}} (\bibinfo{year}{2020}).
\newblock \urlprefix\url{https://doi.org/10.1038/s41598-020-70631-9}.

\bibitem{Balabdaoui2020}
\bibinfo{author}{Balabdaoui, F.} \& \bibinfo{author}{Mohr, D.}
\newblock \bibinfo{title}{Age-stratified discrete compartment model of the
  {COVID}-19 epidemic with application to {S}witzerland}.
\newblock \emph{\bibinfo{journal}{Scientific Reports}}
  \textbf{\bibinfo{volume}{10}} (\bibinfo{year}{2020}).
\newblock \urlprefix\url{https://doi.org/10.1038/s41598-020-77420-4}.

\bibitem{Balcan2009}
\bibinfo{author}{Balcan, D.} \emph{et~al.}
\newblock \bibinfo{title}{Multiscale mobility networks and the spatial
  spreading of infectious diseases}.
\newblock \emph{\bibinfo{journal}{Proceedings of the National Academy of
  Sciences}} \textbf{\bibinfo{volume}{106}}, \bibinfo{pages}{21484--21489}
  (\bibinfo{year}{2009}).
\newblock \urlprefix\url{https://doi.org/10.1073/pnas.0906910106}.

\bibitem{Balcan2010}
\bibinfo{author}{Balcan, D.} \emph{et~al.}
\newblock \bibinfo{title}{Modeling the spatial spread of infectious diseases:
  The {GLobal Epidemic and Mobility} computational model}.
\newblock \emph{\bibinfo{journal}{Journal of Computational Science}}
  \textbf{\bibinfo{volume}{1}}, \bibinfo{pages}{132--145}
  (\bibinfo{year}{2010}).
\newblock \urlprefix\url{https://doi.org/10.1016/j.jocs.2010.07.002}.

\bibitem{Chinazzi2020}
\bibinfo{author}{Chinazzi, M.} \emph{et~al.}
\newblock \bibinfo{title}{The effect of travel restrictions on the spread of
  the 2019 novel coronavirus ({COVID}-19) outbreak}.
\newblock \emph{\bibinfo{journal}{Science}} \textbf{\bibinfo{volume}{368}},
  \bibinfo{pages}{395--400} (\bibinfo{year}{2020}).
\newblock \urlprefix\url{https://doi.org/10.1126/science.aba9757}.

\bibitem{DiDomenico2020}
\bibinfo{author}{Domenico, L.~D.}, \bibinfo{author}{Pullano, G.},
  \bibinfo{author}{Sabbatini, C.~E.}, \bibinfo{author}{Boëlle, P.-Y.} \&
  \bibinfo{author}{Colizza, V.}
\newblock \bibinfo{title}{Impact of lockdown on {COVID}-19 epidemic in
  {{\^{I}}le-de-France} and possible exit strategies}.
\newblock \emph{\bibinfo{journal}{{BMC} Medicine}}
  \textbf{\bibinfo{volume}{18}} (\bibinfo{year}{2020}).
\newblock \urlprefix\url{https://doi.org/10.1186/s12916-020-01698-4}.

\bibitem{Ferguson2020}
\bibinfo{author}{Ferguson, N.} \emph{et~al.}
\newblock \bibinfo{title}{Report 9: Impact of non-pharmaceutical interventions
  {(NPIs)} to reduce {COVID19} mortality and healthcare demand}
  (\bibinfo{year}{2020}).
\newblock \urlprefix\url{http://spiral.imperial.ac.uk/handle/10044/1/77482}.

\bibitem{Newman2002}
\bibinfo{author}{Newman, M. E.~J.}
\newblock \bibinfo{title}{Spread of epidemic disease on networks}.
\newblock \emph{\bibinfo{journal}{Physical Review E}}
  \textbf{\bibinfo{volume}{66}} (\bibinfo{year}{2002}).
\newblock \urlprefix\url{https://doi.org/10.1103/physreve.66.016128}.

\bibitem{Girvan2002}
\bibinfo{author}{Girvan, M.} \& \bibinfo{author}{Newman, M. E.~J.}
\newblock \bibinfo{title}{Community structure in social and biological
  networks}.
\newblock \emph{\bibinfo{journal}{Proceedings of the National Academy of
  Sciences}} \textbf{\bibinfo{volume}{99}}, \bibinfo{pages}{7821--7826}
  (\bibinfo{year}{2002}).
\newblock \urlprefix\url{https://doi.org/10.1073/pnas.122653799}.

\bibitem{arenas2004}
\bibinfo{author}{Arenas, A.}, \bibinfo{author}{Danon, L.},
  \bibinfo{author}{Diaz-Guilera, A.}, \bibinfo{author}{Gleiser, P.~M.} \&
  \bibinfo{author}{Guimera, R.}
\newblock \bibinfo{title}{Community analysis in social networks}.
\newblock \emph{\bibinfo{journal}{The European Physical Journal B}}
  \textbf{\bibinfo{volume}{38}}, \bibinfo{pages}{373--380}
  (\bibinfo{year}{2004}).

\bibitem{Hedayatifar2019}
\bibinfo{author}{Hedayatifar, L.}, \bibinfo{author}{Rigg, R.~A.},
  \bibinfo{author}{Bar-Yam, Y.} \& \bibinfo{author}{Morales, A.~J.}
\newblock \bibinfo{title}{{US} social fragmentation at multiple scales}.
\newblock \emph{\bibinfo{journal}{Journal of The Royal Society Interface}}
  \textbf{\bibinfo{volume}{16}}, \bibinfo{pages}{20190509}
  (\bibinfo{year}{2019}).
\newblock \urlprefix\url{https://doi.org/10.1098/rsif.2019.0509}.

\bibitem{Britton2020}
\bibinfo{author}{Britton, T.}, \bibinfo{author}{Ball, F.} \&
  \bibinfo{author}{Trapman, P.}
\newblock \bibinfo{title}{A mathematical model reveals the influence of
  population heterogeneity on herd immunity to {SARS}-{CoV}-2}.
\newblock \emph{\bibinfo{journal}{Science}} \textbf{\bibinfo{volume}{369}},
  \bibinfo{pages}{846--849} (\bibinfo{year}{2020}).
\newblock \urlprefix\url{https://doi.org/10.1126/science.abc6810}.

\bibitem{Gou2017}
\bibinfo{author}{Gou, W.} \& \bibinfo{author}{Jin, Z.}
\newblock \bibinfo{title}{How heterogeneous susceptibility and recovery rates
  affect the spread of epidemics on networks}.
\newblock \emph{\bibinfo{journal}{Infectious Disease Modelling}}
  \textbf{\bibinfo{volume}{2}}, \bibinfo{pages}{353--367}
  (\bibinfo{year}{2017}).
\newblock \urlprefix\url{https://doi.org/10.1016/j.idm.2017.07.001}.

\bibitem{Gerasimov2021}
\bibinfo{author}{Gerasimov, A.}, \bibinfo{author}{Lebedev, G.},
  \bibinfo{author}{Lebedev, M.} \& \bibinfo{author}{Semenycheva, I.}
\newblock \bibinfo{title}{{COVID}-19 dynamics: A heterogeneous model}.
\newblock \emph{\bibinfo{journal}{Frontiers in Public Health}}
  \textbf{\bibinfo{volume}{8}} (\bibinfo{year}{2021}).
\newblock \urlprefix\url{https://doi.org/10.3389/fpubh.2020.558368}.

\bibitem{Hickson2014}
\bibinfo{author}{Hickson, R.} \& \bibinfo{author}{Roberts, M.}
\newblock \bibinfo{title}{How population heterogeneity in susceptibility and
  infectivity influences epidemic dynamics}.
\newblock \emph{\bibinfo{journal}{Journal of Theoretical Biology}}
  \textbf{\bibinfo{volume}{350}}, \bibinfo{pages}{70--80}
  (\bibinfo{year}{2014}).
\newblock \urlprefix\url{https://doi.org/10.1016/j.jtbi.2014.01.014}.

\bibitem{Dolbeault2021}
\bibinfo{author}{Dolbeault, J.} \& \bibinfo{author}{Turinici, G.}
\newblock \bibinfo{title}{Social heterogeneity and the {COVID}-19 lockdown in a
  multi-group {SEIR} model}.
\newblock \emph{\bibinfo{journal}{Computational and Mathematical Biophysics}}
  \textbf{\bibinfo{volume}{9}}, \bibinfo{pages}{14--21} (\bibinfo{year}{2021}).
\newblock \urlprefix\url{https://doi.org/10.1515/cmb-2020-0115}.

\bibitem{Diekmann1990}
\bibinfo{author}{Diekmann, O.}, \bibinfo{author}{Heesterbeek, J.} \&
  \bibinfo{author}{Metz, J.}
\newblock \bibinfo{title}{On the definition and the computation of the basic
  reproduction ratio {$R_0$} in models for infectious diseases in heterogeneous
  populations}.
\newblock \emph{\bibinfo{journal}{Journal of Mathematical Biology}}
  \textbf{\bibinfo{volume}{28}} (\bibinfo{year}{1990}).
\newblock \urlprefix\url{https://doi.org/10.1007/bf00178324}.

\bibitem{hebert2020}
\bibinfo{author}{H{\'e}bert-Dufresne, L.}, \bibinfo{author}{Althouse, B.~M.},
  \bibinfo{author}{Scarpino, S.~V.} \& \bibinfo{author}{Allard, A.}
\newblock \bibinfo{title}{Beyond {$R_0$}: heterogeneity in secondary infections
  and probabilistic epidemic forecasting}.
\newblock \emph{\bibinfo{journal}{Journal of the Royal Society Interface}}
  \textbf{\bibinfo{volume}{17}}, \bibinfo{pages}{20200393}
  (\bibinfo{year}{2020}).

\bibitem{lloyd2005}
\bibinfo{author}{Lloyd-Smith, J.~O.}, \bibinfo{author}{Schreiber, S.~J.},
  \bibinfo{author}{Kopp, P.~E.} \& \bibinfo{author}{Getz, W.~M.}
\newblock \bibinfo{title}{Superspreading and the effect of individual variation
  on disease emergence}.
\newblock \emph{\bibinfo{journal}{Nature}} \textbf{\bibinfo{volume}{438}},
  \bibinfo{pages}{355--359} (\bibinfo{year}{2005}).

\bibitem{woolhouse1997}
\bibinfo{author}{Woolhouse, M.~E.} \emph{et~al.}
\newblock \bibinfo{title}{Heterogeneities in the transmission of infectious
  agents: implications for the design of control programs}.
\newblock \emph{\bibinfo{journal}{Proceedings of the National Academy of
  Sciences}} \textbf{\bibinfo{volume}{94}}, \bibinfo{pages}{338--342}
  (\bibinfo{year}{1997}).

\bibitem{Driessche2002}
\bibinfo{author}{Van~den Driessche, P.} \& \bibinfo{author}{Watmough, J.}
\newblock \bibinfo{title}{Reproduction numbers and sub-threshold endemic
  equilibria for compartmental models of disease transmission}.
\newblock \emph{\bibinfo{journal}{Mathematical Biosciences}}
  \textbf{\bibinfo{volume}{180}}, \bibinfo{pages}{29--48}
  (\bibinfo{year}{2002}).

\bibitem{Li2021}
\bibinfo{author}{Li, J.}, \bibinfo{author}{Zhong, J.}, \bibinfo{author}{Ji,
  Y.-M.} \& \bibinfo{author}{Yang, F.}
\newblock \bibinfo{title}{A new {SEIAR} model on small-world networks to assess
  the intervention measures in the {COVID}-19 pandemics}.
\newblock \emph{\bibinfo{journal}{Results in Physics}}
  \textbf{\bibinfo{volume}{25}}, \bibinfo{pages}{104283}
  (\bibinfo{year}{2021}).
\newblock
  \urlprefix\url{https://www.sciencedirect.com/science/article/pii/S2211379721004186}.

\bibitem{Allen2008}
\bibinfo{author}{Allen, L. J.~S.}
\newblock \bibinfo{title}{An introduction to stochastic epidemic models}.
\newblock In \bibinfo{editor}{Brauer, F.}, \bibinfo{editor}{van~den Driessche,
  P.} \& \bibinfo{editor}{Wu, J.} (eds.) \emph{\bibinfo{booktitle}{Mathematical
  Epidemiology}}, \bibinfo{pages}{81--130} (\bibinfo{publisher}{Springer-Verlag
  Berlin Heidelberg}, \bibinfo{year}{2008}).
\newblock \urlprefix\url{https://doi.org/10.1007/978-3-540-78911-6_3}.

\bibitem{Althouse2020}
\bibinfo{author}{Althouse, B.~M.} \emph{et~al.}
\newblock \bibinfo{title}{Superspreading events in the transmission dynamics of
  {SARS}-{CoV}-2: Opportunities for interventions and control}.
\newblock \emph{\bibinfo{journal}{{PLOS} Biology}}
  \textbf{\bibinfo{volume}{18}}, \bibinfo{pages}{e3000897}
  (\bibinfo{year}{2020}).
\newblock \urlprefix\url{https://doi.org/10.1371/journal.pbio.3000897}.

\bibitem{TangLu2020}
\bibinfo{author}{Tang, L.} \emph{et~al.}
\newblock \bibinfo{title}{A review of multi-compartment infectious disease
  models}.
\newblock \emph{\bibinfo{journal}{International Statistical Review}}
  \textbf{\bibinfo{volume}{88}}, \bibinfo{pages}{462--513}
  (\bibinfo{year}{2020}).
\newblock
  \urlprefix\url{https://onlinelibrary.wiley.com/doi/abs/10.1111/insr.12402}.

\bibitem{roosa2019}
\bibinfo{author}{Roosa, K.} \& \bibinfo{author}{Chowell, G.}
\newblock \bibinfo{title}{Assessing parameter identifiability in compartmental
  dynamic models using a computational approach: application to infectious
  disease transmission models}.
\newblock \emph{\bibinfo{journal}{Theoretical Biology and Medical Modelling}}
  \textbf{\bibinfo{volume}{16}}, \bibinfo{pages}{1--15} (\bibinfo{year}{2019}).

\bibitem{mehdaoui2021}
\bibinfo{author}{Mehdaoui, M.}
\newblock \bibinfo{title}{A review of commonly used compartmental models in
  epidemiology}.
\newblock \emph{\bibinfo{journal}{arXiv:2110.09642}}  (\bibinfo{year}{2021}).
\newblock \urlprefix\url{https://arxiv.org/abs/2110.09642}.

\bibitem{brauer2008}
\bibinfo{author}{Brauer, F.}
\newblock \bibinfo{title}{Compartmental models in epidemiology}.
\newblock In \emph{\bibinfo{booktitle}{Mathematical epidemiology}},
  \bibinfo{pages}{19--79} (\bibinfo{publisher}{Springer},
  \bibinfo{year}{2008}).

\bibitem{Kenah2008}
\bibinfo{author}{Kenah, E.}, \bibinfo{author}{Lipsitch, M.} \&
  \bibinfo{author}{Robins, J.~M.}
\newblock \bibinfo{title}{Generation interval contraction and epidemic data
  analysis}.
\newblock \emph{\bibinfo{journal}{Mathematical Biosciences}}
  \textbf{\bibinfo{volume}{213}}, \bibinfo{pages}{71--79}
  (\bibinfo{year}{2008}).
\newblock \urlprefix\url{https://doi.org/10.1016/j.mbs.2008.02.007}.

\bibitem{Cirillo2020}
\bibinfo{author}{Cirillo, P.} \& \bibinfo{author}{Taleb, N.~N.}
\newblock \bibinfo{title}{Tail risk of contagious diseases}.
\newblock \emph{\bibinfo{journal}{Nature Physics}}
  \textbf{\bibinfo{volume}{16}}, \bibinfo{pages}{606--613}
  (\bibinfo{year}{2020}).
\newblock \urlprefix\url{https://doi.org/10.1038/s41567-020-0921-x}.

\bibitem{covid19india}
\bibinfo{title}{{COVID19INDIA}}.
\newblock \bibinfo{howpublished}{\url{https://www.covid19india.org/}}.
\newblock \bibinfo{note}{Accessed: 2021-05-27}.

\bibitem{owidcoronavirus}
\bibinfo{author}{Ritchie, H.} \emph{et~al.}
\newblock \bibinfo{title}{Coronavirus pandemic ({COVID-19})}.
\newblock \emph{\bibinfo{journal}{Our World in Data}}  (\bibinfo{year}{2020}).
\newblock \urlprefix\url{https://ourworldindata.org/coronavirus}.

\bibitem{dong2020interactive}
\bibinfo{author}{Dong, E.}, \bibinfo{author}{Du, H.} \&
  \bibinfo{author}{Gardner, L.}
\newblock \bibinfo{title}{An interactive web-based dashboard to track
  {COVID-19} in real time}.
\newblock \emph{\bibinfo{journal}{The Lancet infectious diseases}}
  \textbf{\bibinfo{volume}{20}}, \bibinfo{pages}{533--534}
  (\bibinfo{year}{2020}).

\bibitem{Maltsev2021}
\bibinfo{author}{Maltsev, A.~V.} \& \bibinfo{author}{Stern, M.~D.}
\newblock \bibinfo{title}{Social heterogeneity drives complex patterns of the
  {COVID}-19 pandemic: Insights from a novel stochastic heterogeneous epidemic
  model ({SHEM})}.
\newblock \emph{\bibinfo{journal}{Frontiers in Physics}}
  \textbf{\bibinfo{volume}{8}} (\bibinfo{year}{2021}).
\newblock \urlprefix\url{https://doi.org/10.3389/fphy.2020.609224}.

\end{thebibliography}
\bibliographystyle{naturemag}

\end{document}